\documentclass[a4paper,12pt]{article}
\usepackage{epsfig}
\usepackage{times}
\usepackage{graphicx}
\usepackage{amsmath}
\usepackage{color}
\usepackage{multirow}
\ifx\pdfoutput\undefined
\DeclareGraphicsExtensions{.epsi,.eps}
\else
\DeclareGraphicsExtensions{.pdf}
\fi

\newcommand{\mycite}[1]{\cite{#1}}

\newcommand{\beq}{\begin{equation}}
\newcommand{\eeq}{\end{equation}}
\newcommand{\beqa}{\begin{eqnarray}}
\newcommand{\eeqa}{\end{eqnarray}}
\newcommand{\ket} [1] {\vert #1 \rangle}

\def\ket#1{|\,#1\,\rangle}

\def\opone{\leavevmode\hbox{\small1\kern-3.8pt\normalsize1}}

\begin{document}
	\baselineskip24pt \noindent{\Large\bf Experimental Quantum Multiparty Communication Protocols}\\[3mm]
	\noindent {\small \bf Massimiliano Smania$^{1,\star}$, Ashraf M. Elhassan$^{1,\star}$, Armin Tavakoli$^{1}$ \& Mohamed Bourennane$^{1,\ddag}$}\\
	\noindent {\it $^{1}$Physics department, Stockholm University, S-10691, Stockholm, Sweden}\\[2mm]
	\noindent {\it $^{*}$These authors contributed equally to this work}\\[2mm]
\noindent
\textbf{
 Quantum information science breaks limitations of conventional information transfer, cryptography and computation by using quantum superpositions or entanglement as resources for information processing. Here, we report on the experimental realization of three-party quantum communication protocols using  single  three-level quantum system  (qutrit) communication: secret sharing, detectable Byzantine agreement, and communication complexity reduction for a  three-valued function.  We have implemented these three schemes using the same optical fiber interferometric setup. Our  realization is easily scalable without sacrificing detection efficiency or generating extremely complex many-particle entangled states.}

%\maketitle
\newpage
{\bf Introduction}

Many tasks in communications, computation, and cryptography can be enhanced beyond classical limitations by using quantum resources. Such quantum technologies often rely on distributing strongly correlated data that cannot be reproduced with classical theory i.e. it violates a Bell inequality\mycite{Bell64}. To violate a Bell inequality, the parties involved in the scheme must share an entangled quantum state on which they perform suitable local measurements returning outcomes that can be locally processed and communicated by classical means. Such entanglement-assisted schemes  have been shown successful in a wide variety of information processing tasks, including secret sharing for which additional security features are enabled, detectable Byzantine Agreement for which a classically unsolvable task can be solved, and reduction of communication complexity for which optimal classical techniques are outperformed. Let us shortly introduce these three communication protocols. 

Secret sharing is a cryptographic primitive that can conceptually be regarded as a generalization of quantum key distribution\mycite{BB84, E91}. Secret sharing schemes have wide applications in secure multiparty computation and management of keys in cryptography. In such schemes, a message (secret) is divided in shares distributed to recipient parties in such a way that some number of parties must collaborate in order to reconstruct the message. However, the security of classical secret sharing relies on limiting assumptions of the computation power available to an adversary. Quantum cryptography introduces the concept of unconditional security, and can improve security beyond classical constraints. Quantum secret sharing protocols have been proposed with parties sharing a multipartite qubit entangled state\mycite{ZZWH98,HBB99} where their security is linked to Bell inequality violations.

A fundamental problem in fault-tolerant distributed computing is to achieve coordination between computer processes in spite of some processes randomly failing due to e.g. crashing, transmission failure or distribution of incorrect information in the network. For example, such coordination applies to the problem of synchronizing the clocks of individual processes in distributed networks,  which is pivotal in many technologies including data transfer networks and telecommunication networks. A method to achieve synchronization is to use interactive consistency algorithms in which all nonfaulty processes reach a mutual agreement about all the clocks\mycite{LM85}. Interactive consistency is achieved through solving the problem of Byzantine agreement, which can be solved only if less than one-third of the processes are faulty\mycite{LSP82}. However, for most applications, it is sufficient to consider a scenario called detectable Byzantine agreement (DBA), where the processes either achieve mutual agreement or jointly exit the protocol. Several quantum protocols based on multipartite entanglement have been proposed for achieving the DBA even in the presence of one-third or more faulty processes, thus breaking the classical limitation\mycite{FGM01,C03,GBKCW07}.

In communication complexity problems (CCPs), separated parties performing local computations exchange information in order to accomplish a globally defined task, which is impossible to solve single-handedly. Here, we consider the situation in which one would like to maximize the probability of successfully solving a task with a restricted amount of communication\mycite{Y79}. Such
studies aim, for example, at speeding-up a distributed computation by increasing the communication efficiency, or at optimizing VLSI circuits and data structures\mycite{KN97}. Quantum protocols involving multipartite entangled states have been shown to be superior to classical protocols for a number of CCPs\mycite{RB97,BZZ02}.

Quantum multiparty communication protocols that require only sequential communication of single qubits and no shared multipartite entanglement have been proposed for secret sharing {\mycite{STBKZW05}} and CCP \mycite{TSBBZW05}, and CCP  using the quantum Zeno effect \mycite{TAHB15}. Very recently generalizations to d-level quantum system (called a qudit) have been proposed. These  protocols are multiparty quantum secret sharing\mycite{THZB15} and  a quantum solution to the DBA, which can then be used to achieve clock synchronization in the presence of an arbitrary number of faulty processes by efficient classical means of communications\mycite{TCZB15}. Beside experimentally realizing these protocols, we propose and demonstrate a new single qudit protocol for a multiparty CCP, that outperforms any classical counterpart.

Although the mentioned information processing tasks cover very different topics; cryptography, synchronization, and communication complexity, we will show that the quantum schemes that distribute these correlated data sets uphold strong similarities and the differences emerge from the classical processing of the correlated data required to execute these protocols. 

Our single qudit communication protocols hold several experimental advantages in scalability over the corresponding entanglement-assisted schemes. While entanglement-assisted protocols typically require the preparation of a high fidelity $N$-partite $d$-level entangled quantum state, the single qudit protocols earn their name from requiring only the preparation of a single qudit independently of the number of parties, $N$, involved in the protocol. Furthermore, in the likely case of parties using non-ideal detectors with efficiencies $\eta\in[0,1]$, entanglement-assisted protocols require $N$ detections and therefore succeed with an exponentially decreasing probability, approximately $\eta^N$, while single qudit protocols only require a single detection which succeeds with probability $\eta$, independently of $N$.

{\bf Communication protocols} 

In this report, we will present for the first time the experimental realization of quantum communication protocols, secret sharing, DBA and clock synchronization, and reduction of communication complexity in a multipartite setting involving three parties, Alice, Bob and Charlie, communicating three-level quantum states (qutrits). We will now very briefly present these  protocols.

{\bf Secret Sharing}

\begin{figure}[h]
	\begin{center}
		\includegraphics[width=0.8\columnwidth]{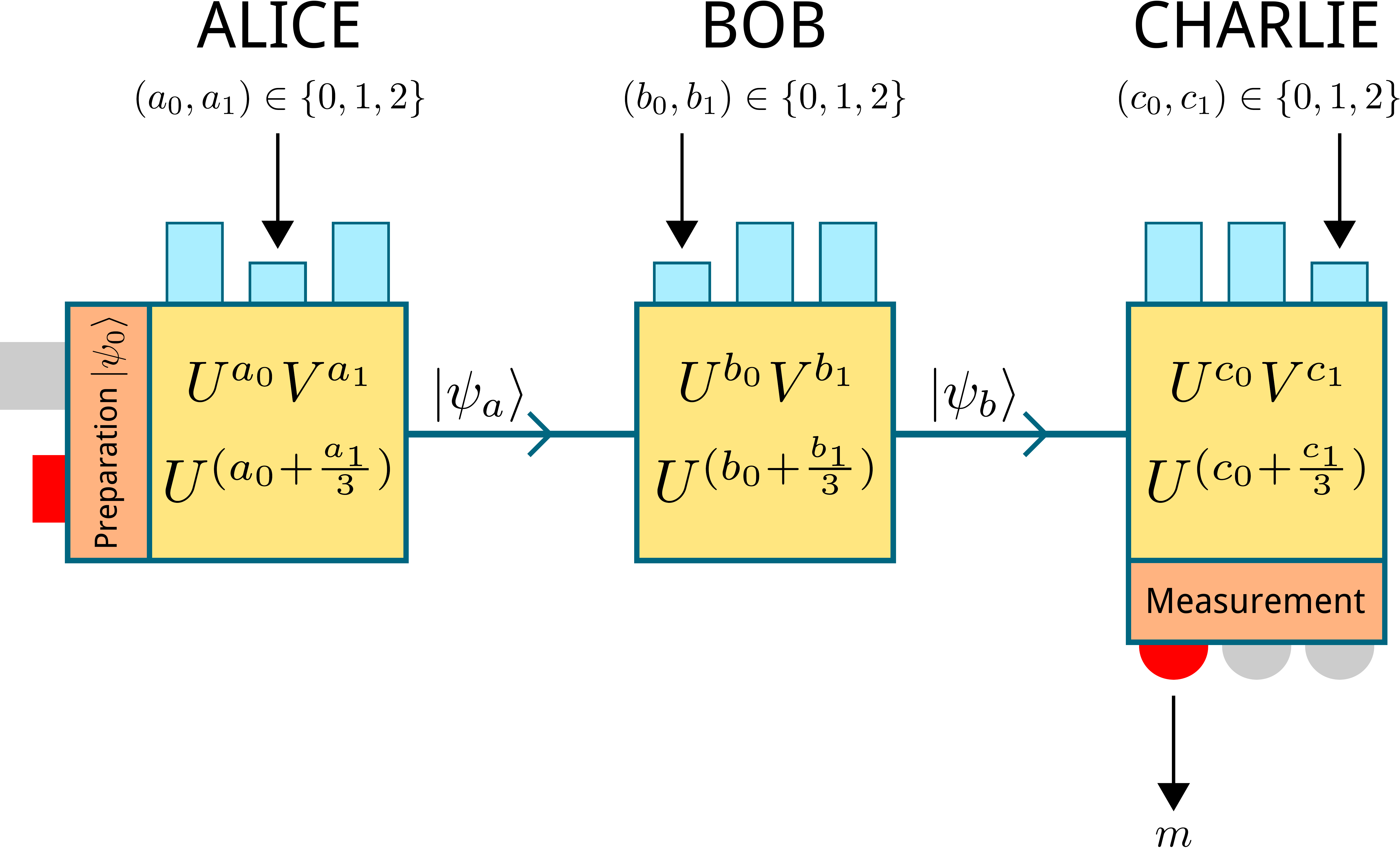}
		\caption{Schematic illustration for the three single qutrit three-party communication protocols.
		Alice prepares  state $|\psi\rangle$, Alice, Bob, and Charlie act with the operation  $U^{i}V^{j}$ sequentially on the received state according to their input data $(i,j)$, where $(i,j)$ are $(a_0,a_1)$
		$(b_0,b_1)$, and $(c_0,c_1)$ for Alice, Bob, and Charlie respectively. Finally, Charlie performs a measurement on the qutrit in the Fourier basis.}\label{schema}
	\end{center}
\end{figure}

Alice (a.k.a. the \textit{distributor}) prepares the initial qutrit state $|\psi\rangle=\frac{1}{\sqrt{3}}\left(|0\rangle+|1\rangle+|2\rangle\right)$ and applies her action $U^{a_0}V^{a_1}$ on the state $|\psi\rangle$ according to her input data $(a_0,a_1)$, where $a_0$ and $a_1$ are two pseudo-random independent numbers in the set $\{0,1,2\}$, and operators $U$ and $V$ are given by
\begin{eqnarray}
U=|0\rangle\langle 0|+e^{\frac{2\pi i}{3}} |1\rangle\langle 1|+e^{-\frac{2\pi i}{3}}|2\rangle\langle 2|	 \label{eq:V}\\
V=|0\rangle\langle 0|+e^{\frac{2\pi i}{3}} |1\rangle\langle 1|+e^{\frac{2\pi i}{3}}|2\rangle\langle 2| . 	 \label{eq:H}
\end{eqnarray}
Then she sends the qutrit to Bob, who according to his input data $(b_0,b_1)$ acts on the qutrit with operator $U^{b_0}V^{b_1}$, and sends the state to Charlie who acts on the qutrit with operator $U^{c_0}V^{c_1}$, where $(c_0,c_1)$ are his input data.
Finally, Charlie performs a measurement on the qutrit in the Fourier basis $\left\lbrace\frac{1}{\sqrt{3}}(1,1,1),\frac{1}{\sqrt{3}}(1,e^{\frac{2\pi i}{3}},e^{-\frac{2\pi i}{3}}),\frac{1}{\sqrt{3}}(1,e^{-\frac{2\pi i}{3}},e^{\frac{2\pi i}{3}})\right\rbrace$, obtaining a trit outcome $m$ (see Fig. \ref{schema}).
In random order, the parties then announce their data $a_1,\ b_1,\ c_1$, and if condition $a_1+b_1+c_1=0\mod{3}$ is verified, the round is treated as valid and equation $a_0+b_0+c_0=m\mod{3}$ produces the shared secret. Otherwise if $a_1+b_1+c_1 \neq 0\mod{3}$ the qutrit is not in an eigenstate of the measurement operator at the time of measurement. Thus, the outcome $m$ is random and the run is discarded. At this point all users should publicly announce $a_0$, $b_0$ and $c_0$ for a relevant number of runs and estimate the quantum trit error rate (QTER) defined as $QTER=\textit{number of incorrect outcomes}/\textit{total number of outcomes}$. Finally, to reconstruct the shared secret at least two users are required to collaborate \mycite{THZB15}.

%%%%%%%%%%%%%%%%%%%%%%%%%%%%%%%%%%% NEW
Secret sharing schemes can be subjected not only to eavesdropping attacks but also to attacks from parties within the scheme. Examples are known in which such attacks can breach the security of secret sharing schemes \cite{H07}. In the supplementary material we outline a scheme enforcing security  that can, at the cost of a lower efficiency, arbitrarily minimize the impact of such attacks. Furthermore, we  mention that the full security can be obtained from device independent implementations of entanglement based quantum key distribution \mycite{ABGMPS07}. However, to our knowledge, there are no device-independent protocols for secret sharing. In our proposed protocol we assume that the users have control over the devices. 
%%%%%%%%%%%%%%%%%%%%%%%%%%%%%%%%%%%

{\bf Detectable Byzantine Agreement} 

In order to solve the DBA problem, the three processes (i.e. parties) need to share data in the form of lists $l_k$ of numbers subject to specific correlations, and the distribution must be such that the list $l_k$ held by process $P_k$ is known only to $P_k$ where $k = 1,2,3$. Quantum mechanics provides methods to generate and securely distribute such data. In this case, Alice's state preparation and each user's action are the same as in the previous protocol, except for $b_0$ and $c_0$ being bits instead of trits. The difference is in the data processing part: if the measurement outcome is ``0'', the parties reveal $a_1,b_1,c_1$ and if condition $a_1+b_1+c_1=0\mod{3}$ is satisfied, the round is treated as valid. It follows that they now hold one of the data sets $(a_0,b_0,c_0)\in\{(0,0,0),(1,1,1),(2,1,0),(2,0,1)\}$ from which the DBA can be solved\cite{TCZB15}.

{\bf Communication Complexity Reduction}

In the single qutrit protocol for reducing communication complexity, the distributor supplies Alice, Bob and Charlie with two pseudo-random trits each, $(a_0,a_1)$, $(b_0,b_1)$ and $(c_0,c_1)$. Each party's pair can be mapped into an integer by defining $S_x\equiv 3x_0+x_1 \in \{0,...,8\}$, with $x\in\{a,b,c\}$. The distributor promises the parties that $S_a+S_b+S_c=0\mod{3}$ and asks Charlie to guess the value of function $T=(S_a+S_b+S_c\mod{9})/3$ given that only two (qu)trits may be communicated in total.
After the  $|\psi\rangle$ state preparation, Alice acts with $U^{\frac{S_a}{3}}$ (with $U$ defined as in Eq. \eqref{eq:V}) and sends the qutrit to Bob, who applies $U^{\frac{S_b}{3}}$ before forwarding it to Charlie.
Finally, after applying $U^{\frac{S_c}{3}}$, Charlie performs a measurement on the resulting state $|\psi_{final}\rangle=\frac{1}{\sqrt{3}}(|0\rangle+e^{\frac{2\pi i}{3}T}|1\rangle+e^{-\frac{2\pi i}{3}T}|2\rangle)$.
This state is an element of the Fourier basis, so a measurement in this basis will output the correct value of function $T$ with (ideally) unit probability.

The CCP is to maximize the success probability of guessing $T$ correctly, with the given communication restrictions. We have just seen that this success probability, save for experimental errors, is $100\%$ with our quantum protocol. However, it can be shown (see supplementary material) that the optimal classical protocol  achieves only a success probability of $7/9\approx 0.778$, which is clearly inferior to that of the quantum protocol.

{\bf Experimental Realization}

\begin{figure}[h]
	\begin{center}
		\includegraphics[width=1\columnwidth]{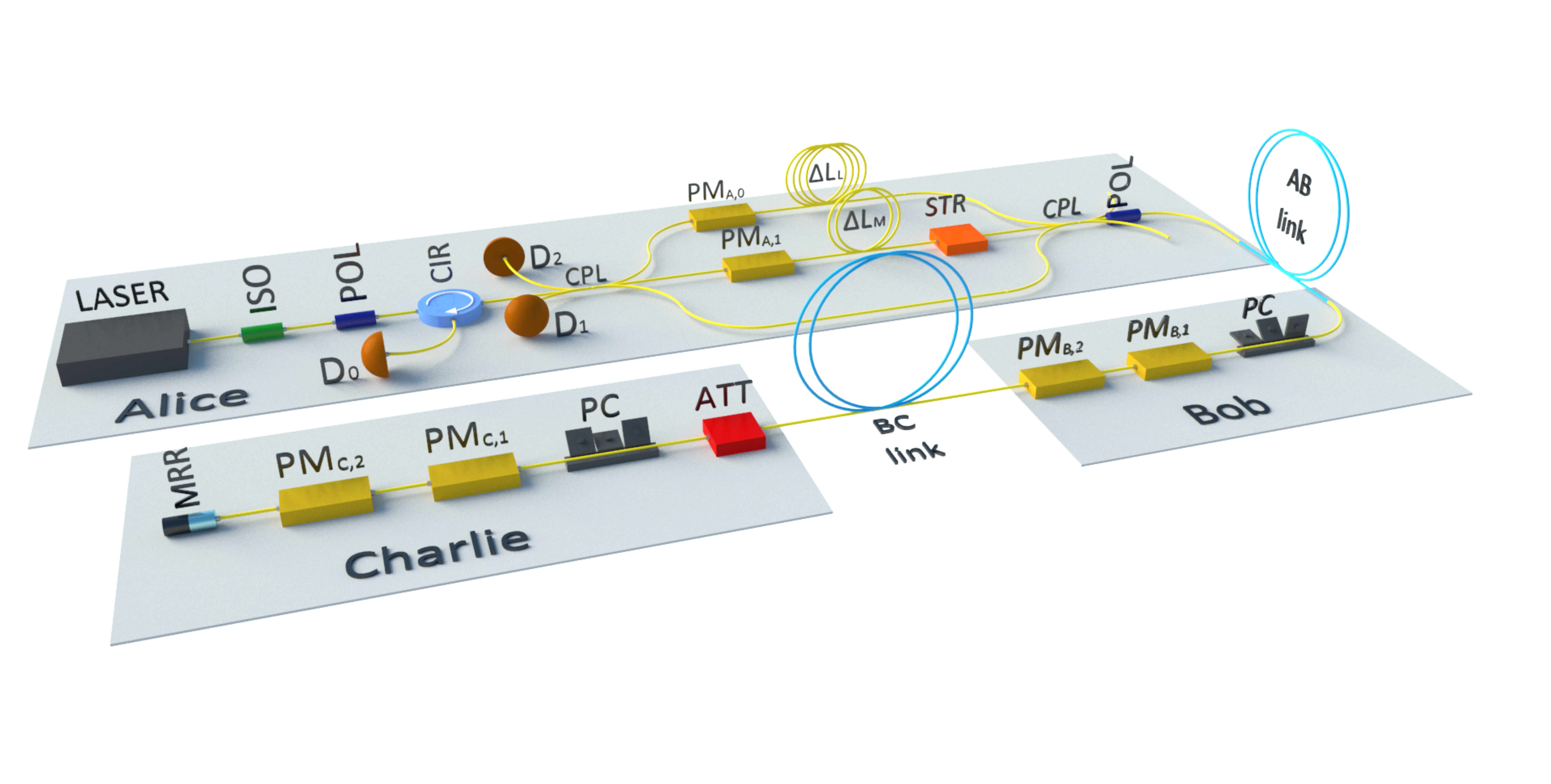}
		\caption{Experimental setup used in this work. The components are: isolator (ISO), polarizers (POL), circulator (CIR), two 3x3 fiber couplers (CPL), phase modulators (PM), fiber stretcher (STR), polarization controllers (PC), variable digital attenuator (ATT), retro-reflective mirror (MRR) and three single photon avalanche detectors (D$_0$, D$_1$, D$_2$). The three parties' stations are polarization maintaining, while the links connecting them are single mode fibers.}\label{setup}
	\end{center}
\end{figure}

We have realized the three above-mentioned communication protocols with the same optical setup reported in Fig. \ref{setup}.
The setup is based on a three-arm Mach-Zehnder-like interferometer built with optical fibers and a retro-reflective mirror (this configuration is a practical solution to the natural phase-drift problem which affects every Mach-Zehnder interferometer, further complicated here by the fact that we have three paths).
The information transmitted between users is encoded in relative phase differences between the three states constituting the qutrit.
The state preparation is carried out by sending light pulses from a $1550$ nm diode laser (ID300 by ID Quantique) to the first 3x3 coupler of the Mach-Zehnder interferometer. The laser repetition rate is  $100$ kHz.
The outcome after the second coupler is a superposition of the three paths, so that the optical phase of each pulse of the qutrit can be individually modulated with commercial phase modulators (COVEGA Mach-10 Lithium Niobate Modulators).
The delays in the interferometer are $\Delta L_M=68.40 \pm 0.05$ ns and $\Delta L_L=136.80 \pm 0.05$ ns.
On the way to the mirror, users passively let pass the qutrit through while after the reflection Charlie, Bob and Alice sequentially act on the qutrit with a combination of operators $U$ and $V$ (see Eqs. \eqref{eq:V} and \eqref{eq:H}).
After passing through the three arms on their way back, the three pulses recombine at the first coupler and depending on their relative phases, yield different interference counts at the single photon detectors (Princeton Lightwaves PGA600). These gated detectors provide $20\%$ quantum efficiency and approximately $10^{-5}$ dark count probability. Importantly, in order to prevent possible eavesdropping attacks, each pulse is attenuated to single photon level by a digital variable attenuator (OZ Optics DA-100) at Charlie's station output.
We would like to emphasize that phase modulators are polarization sensitive, and for this reason they include a horizontal polarizer at the output port. Therefore, controlling polarization throughout the setup is crucial. We thus choose to use polarization maintaining fiber components for all three parties' stations. However, in order to make the configuration more realistic, links between users are standard single mode fibers. Therefore, polarization controllers have been placed after these fiber links. Finally, the whole experiment was controlled by an FPGA card that worked both as master clock and trigger source, for the electronics driving laser and phase modulators, and for the single photon detectors.
For future and practical implementations of these communication protocols, one needs to use a bright true or heralded single photon source, integrated optics interferometer, and high quantum efficiency superconducting single photon detectors.

{\bf Results}

Each protocol setting was run $100000$ times per second (i.e. $10^5$ laser triggers) and the collected data was used to calculate the QTER. Due to the substantial loss from the setup itself (mainly in the phase modulators) and to the $20\%$ detection efficiency, the final amount of runs with detection was  $400$ per setting.
Our results for secret sharing and DBA experiments are reported in Tab. \ref{tab1} and in Tab. \ref{tab2} respectively.
We can easily see that QTER for the secret sharing and DBA protocols are always below $10\%$. Our results are better than other results obtained with entanglement-based two-party quantum key distribution protocols \mycite{zei_qutrits}, and QTER clearly are below the $15.95\%$ security threshold of qutrit based quantum key distribution \mycite{CBKN01}. Therefore secure communication can be obtained with this configuration \mycite{THZB15}.
Consistently, CCP experimental results, of which a sample of obtained data is reported in Tab. \ref{tab3}, show success probabilities always above $90\%$, therefore proving the superiority of the quantum protocol to any classical protocol (limited to $77.8$\% success probability).

The primary source of QTER is the so called ``dark counts''. Our detectors' average dark count probabilities, measured with $10^6$ runs, are $5.9\cdot 10^{-5}$, $2.8\cdot 10^{-5}$ and $20.5\cdot 10^{-5}$ per trigger for detectors 0, 1 and 2 respectively. Considering our measurements, these dark counts contribute up to half of the QTER. Other important systematic contributions to the QTER are due to the phase drift affecting the interferometer. This phase drift causes two problems: it slightly changes the relative phases from the desired settings and it forces a recalibration of the phases before each experiment. Both these contributions can be quantified, by propagating phase errors in interference equations (see supplementary material), to be approximately $1\%$ each to QTERs.

\begin{table}[t]
	\begin{center}
		\begin{tabular}{cc|cc|cc|c|ccc|c}
			\multicolumn{2}{c|}{Alice} & \multicolumn{2}{c|}{Bob} & \multicolumn{2}{c|}{Charlie} & \multirow{2}{*}{$m$} & \multicolumn{3}{c|}{\textbf{Counts}} & QTER \\ \cline{1-6} \cline{8-10}
			$a_0$ & $a_1$ & $b_0$ & $b_1$ & $ c_0 $ & $c_1$ & & D$_0$ \vline & D$_1$ \vline & D$_2$ & [\%] \\ \hline
			0 & 0 & 0 & 0 & 2 & 0 & 2 & 7 & 5 & \textbf{210} & 5.41 \\
			1 & 0 & 0 & 0 & 1 & 0 & 2 & 7 & 6 & \textbf{261} & 4.74 \\
			2 & 0 & 2 & 1 & 2 & 2 & 0 & \textbf{375} & 15 & 26 & 9.86 \\
			0 & 1 & 2 & 2 & 1 & 0 & 0 & \textbf{391} & 10 & 29 & 9.07 \\
			1 & 1 & 0 & 1 & 2 & 1 & 0 & \textbf{336} & 7 & 23 & 8.20 \\		
			2 & 1 & 1 & 1 & 1 & 1 & 1 & 7 & \textbf{373} & 22 & 7.21 \\
			0 & 2 & 2 & 0 & 0 & 1 & 2 & 16 & 13 & \textbf{313} & 8.48 \\
			1 & 2 & 2 & 2 & 2 & 2 & 2 & 19 & 8 & \textbf{248} & 9.82 \\
			2 & 2 & 1 & 0 & 1 & 1 & 1 & 9 & \textbf{284} & 22 & 9.84 \\ \hline
			1 & 0 & 0 & 2 & 2 & 0 & random & 102 & 98 & 94 & 65.31 \\
			2 & 2 & 0 & 0 & 0 & 0 & random & 89 & 75 & 71 & 62.13 \\
		\end{tabular}
		\caption{Results for the Secret Sharing protocol. \label{tab1}}
	\end{center}
\end{table}
\newpage
\begin{table}[t]
	\begin{center}
		\begin{tabular}{c c|c c|c c|c|ccc|c}
			\multicolumn{2}{c|}{Alice} & \multicolumn{2}{c|}{Bob} & \multicolumn{2}{c|}{Charlie} & \multirow{2}{*}{$m$} & \multicolumn{3}{c|}{\textbf{Counts}} & QTER \\ \cline{1-6} \cline{8-10}
			$a_0$ & $a_1$ & $b_0$ & $b_1$ & $ c_0 $ & $c_1$ & & D$_0$ \vline & D$_1$ \vline & D$_2$ & [\%] \\ \hline
			0 & 0 & 1 & 0 & 1 & 0 & 2 & 16 & 11 & \textbf{337} & 7.42 \\
			1 & 0 & 0 & 0 & 0 & 0 & 1 & 16 & \textbf{320} & 19 & 9.86 \\
			\textbf{2} & 0 & \textbf{1} & 0 & \textbf{0} & 0 & 0 & \textbf{347} & 13 & 20 & 8.68 \\
			\textbf{0} & 1 & \textbf{0} & 1 & \textbf{0} & 1 & 0 & \textbf{363} & 13 & 20 & 8.33 \\
			1 & 1 & 1 & 1 & 0 & 1 & 2 & 11 & 17 & \textbf{333} & 7.76 \\		
			\textbf{2} & 1 & \textbf{0} & 1 & \textbf{1} & 1 & 0 & \textbf{309} & 9 & 13 & 6.65 \\
			0 & 2 & 1 & 1 & 0 & 0 & 1 & 7 & \textbf{277} & 19 & 8.58 \\
			1 & 2 & 0 & 2 & 1 & 2 & 2 & 9 & 18 & \textbf{274} & 8.97 \\
			\textbf{2} & 2 & \textbf{1} & 2 & \textbf{0} & 2 & 0 & \textbf{300} & 7 & 26 & 9.91
		\end{tabular}
		\caption{Results for the DBA protocol. \label{tab2}}
	\end{center}
\end{table}
\newpage

\begin{table}[t]
	\begin{center}
		\begin{tabular}{c|c|c|c|ccc|c}
			Alice & Bob & Charlie & \multirow{2}{*}{ T } & \multicolumn{3}{c|}{\textbf{Counts}} & SP \\ \cline{5-7}
			$S_a$ & $S_b$ & $S_c$ & & D$_0$ \vline & D$_1$ \vline & D$_2$ & [\%] \\ \hline
			0 & 1 & 8 & 0 & \textbf{350} & 7 & 28 & 90.91 \\
			0 & 2 & 1 & 1 & 8 & \textbf{284} & 23 &  92.53 \\
			1 & 5 & 0 & 2 & 14 & 14 & \textbf{255} &  90.11 \\
			1 & 6 & 2 & 0 & \textbf{337} & 5 & 29 &  90.84 \\
			2 & 7 & 3 & 1 & 13 & \textbf{268} & 16 &  90.24 \\
			2 & 0 & 4 & 2 & 10 & 2 & \textbf{204} &  94.44 \\
			3 & 2 & 4 & 0 & \textbf{302} & 8 & 22 &  90.96 \\
			3 & 1 & 8 & 1 & 8 & \textbf{358} & 22 &  92.27 \\
			4 & 8 & 3 & 2 & 10 & 13 & \textbf{269} &  92.12 \\
			4 & 5 & 0 & 0 & \textbf{332} & 12 & 21 &  90.96 \\
			5 & 6 & 1 & 1 & 21 & \textbf{370} & 19 &  90.24 \\
			5 & 4 & 6 & 2 & 14 & 18 & \textbf{297} &  90.27 \\
			6 & 2 & 1 & 0 & \textbf{298} & 3 & 28 &  90.30 \\
			6 & 8 & 7 & 1 & 6 & \textbf{297} & 18 &  92.52 \\
			7 & 3 & 5 & 2 & 6 & 13 & \textbf{232} &  92.43 \\
			7 & 0 & 2 & 0 & \textbf{264} & 12 & 12 &  91.67 \\
			8 & 2 & 2 & 1 & 7 & \textbf{385} & 31 &  90.40 \\
			8 & 8 & 8 & 2 & 13 & 11 & \textbf{229} &  90.51 \\
		\end{tabular}
		\caption{Results for the  communication complexity reduction protocol. \label{tab3}}
	\end{center}
\end{table}

{\bf Conclusion}

We have experimentally realized three-party quantum communication protocols using single qutrit communication: secret sharing, detectable Byzantine agreement, and communication complexity reduction for a  three-valued function.  We have implemented for the fist time these three protocols  using the same optical fiber interferometric setup.
Our novel protocols are based on single quantum system communication rather than entanglement. Moreover, the number of detectors (detector noise) used in our schemes is independent of the number of parties participating in the protocol. Our  realization is easily scalable without sacrificing detection efficiency or generating extremely complex many-particle entangled states. These breakthrough and advances  make multiparty communication tasks feasible. They become technologically comparable to quantum key distribution, so far the only commercial application of quantum information.
Finally, our methods and techniques can be generalized to other communication protocols. These protocols can also be easily adapted for other encodings and physical systems.

%%%%%%%%%%%%%%%%%%%%%%%%%%%%%%%%%%%%%%%%%%%%%%%%%%%%%%%%%%%%%%%%%
% Supplementary information
%%%%%%%%%%%%%%%%%%%%%%%%%%%%%%%%%%%%%%%%%%%%%%%%%%%%%%%%%%%%%%%%%

\textbf{Supplementary Information} is linked to the online version of the paper at www.nature.com/nature.\\

%%%%%%%%%%%%%%%%%%%%%%%%%%%%%%%%%%%%%%%%%%%%%%%%%%%%%%%%%%%%%%%%%
% Acknowledgements
%%%%%%%%%%%%%%%%%%%%%%%%%%%%%%%%%%%%%%%%%%%%%%%%%%%%%%%%%%%%%%%%%

{\bf Acknowledgements} This work was supported by the Swedish Research Council.\\

%%%%%%%%%%%%%%%%%%%%%%%%%%%%%%%%%%%%%%%%%%%%%%%%%%%%%%%%%%%%%%%%%
% Competing interests statement
%%%%%%%%%%%%%%%%%%%%%%%%%%%%%%%%%%%%%%%%%%%%%%%%%%%%%%%%%%%%%%%%%
{\textbf{Author contribution} M.B. initiated and proposed the project. M.S .and A.M.E. design, performed the experiment, and analyzed  the data. A.T. carried out the theoretical calculation for the CCP and scheme enforcing security of secret sharing protocols. All the authors discussed the results and wrote the manuscript.\\

{\bf Competing interests statement} The authors declare that they
have no competing financial interests.\\
%%%%%%%%%%%%%%%%%%%%%%%%%%%%%%%%%%%%%%%%%%%%%%%%%%%%%%%%%%%%%%%%%
% Correspondence
%%%%%%%%%%%%%%%%%%%%%%%%%%%%%%%%%%%%%%%%%%%%%%%%%%%%%%%%%%%%%%%%%

{\bf Correspondence} and requests for materials should be addressed
to M.B. (e-mail: boure@fysik.su.se).
%%%%%%%%%%%%%%%%%%%%%%%%%%%%%%%%%%%%%%%%%%%%%%%%%%%%%%%%%%%%%%%%%

\newpage
	\baselineskip24pt \noindent{\Large\bf Experimental Quantum Multiparty Communication Protocols\\ Supplementary Material}\\[3mm]
	\noindent {\small \bf Massimiliano Smania$^{1,\star}$, Ashraf M. Elhassan$^{1,\star}$, Armin Tavakoli$^{1}$ \& Mohamed Bourennane$^{1,\ddag}$}\\
	\noindent {\it $^{1}$Physics department, Stockholm University, S-10691, Stockholm, Sweden}\\[2mm]
	\noindent {\it $^{*}$These authors contributed equally to this work}\\[2mm]
	\noindent
	
	%%%%%% NEW SECTION
	\section{Scheme enforcing security}
	In secret sharing schemes one may face attacks from cheating parties within the scheme. Since in a three party scheme, it is not meaningful to have more than one cheater, let us assume that the cheating party is Bob who attempts to access Alice's secret. Since Bob is involved in the scheme, he may employ a sophisticated strategy based on e.g. a quantum memory and an entangled ancilla state which can allow him to pass the standard security check undetected. The success of any such strategy must in some manner depend on Bob gaining information on some local data ($a_0, c_0$) of Alice and Charlie. However, as the three parties announce their data in random order Bob cannot always succeed with his cheating, and thus we can upper bound the probability of Bob successfully cheating in a single round by $p_{cheat}\leq 2/3$. Here, we will outline a privacy amplification scheme that allows Alice and Charlie to arbitrarily minimize Bob's chances of inferring the secret. 
	
	The standard protocol is repeated many times and some rounds are used to estimate the QTER. Assuming the QTER does not imply any security breach, this leaves $L$ successful rounds. We let $a_0^{(l)}$, $b_0^{(l)}$, $c_0^{(l)}$ be the private data of Alice, Bob and Charlie respectively in the $l$'th such round, and we let $m^{(l)}$ be the measurement outcome of Charlie in the round. All parties then map their $L$ trits into a single trit by setting $x'_0\equiv \sum_{l=1}^{L}x_0^{(l)}-m^{(l)}\delta_{x,a} \mod{3}$ with $x\in\{a,b,c\}$. The new data now satisfies $a'_0 + b'_0+c'_0 = 0\mod{3}$. Unless Bob successfully cheats in all $L$ rounds, which happens with a probability $\bar{p}_{cheat}=p_{cheat}^L$, he can infer no knowledge on the private data of Alice or Charlie. Thus, the number of rounds that warrants the cheating success probability to be no higher than $\bar{p}_{cheat}$ is 
	$L=\lceil \log{\bar{p}_{cheat}}/\log{p_{cheat}}\rceil$ i.e. at the cost of repeating the protocol several times, the cheating probability can be arbitrarily minimized. If we consider an example of a known (G. P. He, {\it Phys. Rev. Lett.} 2007; {\bf 98}: 028901-1.) such cheating attack in a three party secret sharing protocol in which Bob was found to have cheating success probability per round of $p_{cheat}=1/3$, and we require that $\bar{p}_{cheat}\leq 10^{-4}$, then we find that we need $L=9$ rounds to achieve this level of security.    
	
	\section{Communication Complexity Reduction: classical bound }
	The quantum protocol for communication complexity reduction presented in our paper uses a single qutrit and achieves the ideal success probability of $100\%$. We will now prove that our quantum protocol is superior to any classical one. \\
	To find the best classical protocol, we can without loss of generality consider only deterministic strategies and disregards mixtures of such. Alice has to send the value of some three-valued function $f_A\in\{1,e^{\frac{2\pi i}{3}},e^{-\frac{2\pi i}{3}}\}$ to Bob, and similarly, Bob has to send the value of some three-valued function $f_B$ to Charlie, who outputs a final guess of $T$ determined by a three-valued function $f_C$. All such functions take the form 
	\begin{equation}
	f(x)=q_0+q_1e^{\frac{2\pi i}{3}x}+q_2e^{-\frac{2\pi i}{3}x}
	\end{equation}
	with suitable choices of coefficients $q_0,q_1,q_2$.
	Therefore, we write 
	\begin{equation}
	f_C=q_0^{(C)}(c_1,f_A,f_B)+q_1^{(C)}(c_1,f_A,f_B)e^{\frac{2\pi i}{3}c_0}+q_2^{(C)}(c_1,f_A,f_B)e^{-\frac{2\pi i}{3}c_0}, 
	\end{equation}
	where we have emphasized the dependence of the coefficients on the received data $f_A,f_B$ from subsequent parties.
	However, we must have $q_0^{(C)}=0$ since otherwise, the strategy is randomized. It is straightforward to find that this implies that only one of the coefficients $q_1^{(C)},q_2^{(C)}$ can be non-zero, and that its value must be either $1,e^{\frac{2\pi i}{3}}$ or $e^{-\frac{2\pi i}{3}}$.
	Therefore, we find that 
	\begin{equation}
	f_C=q_{r_A}^{(A)}q_{r_B}^{(B)}q_{r_C}^{(C)}e^{\frac{2\pi i}{3}(r_Aa_0+r_Bb_0+r_Cc_0)} 
	\end{equation}
	for some set $r_A,r_B,r_C$ subject to $r_A,r_B,r_C\in\{1,2\}$. Evidently, we should choose $r_A=r_B=r_C=1$ such that the contribution from $a_0,b_0,c_0$ to the task function is correct. Then, since $a_1,b_1,c_1$ are uniformly distributed and we are promised that $a_1+b_1+c_1=0\mod{3}$, there are nine different sets $(a_1,b_1,c_1)$ of which only $(0,0,0)$ sums to $0$, only $(2,2,2)$ sums to $6$ and the remaining seven sets sum to $3$. Therefore, let $q_1^{(A)}=q_1^{(B)}=1$ and $q_1^{(C)}=e^{\frac{2\pi i}{3}}$ which yields
	\begin{equation}
	f_C=e^{\frac{2\pi i}{3}(a_0+b_0+c_0+1)}.
	\end{equation} 
	In conclusion, the optimal classical success probability is $P(f_C=T)=7/9\approx 0.778$, which is clearly inferior to the unitary one for the quantum protocol.

	\section{Encoding Settings}
	In Tables S1 and S2,  we report encoding settings used by Alice, Bob and Charlie for the three experiments. We write directly the angular phase shifts to be applied to the three state qutrits exchanged among the users.
	
	\begin{table}[h!]
		\begin{center}
			\begin{tabular}{cc|ccc|ccc}
				\multicolumn{8}{c}{\textbf{Secret sharing \& DBA}} \\ \hline
				\multicolumn{2}{c|}{\textbf{Settings}} & \multicolumn{3}{c|}{\textbf{Alice}} & \multicolumn{3}{c}{\textbf{Bob \& Charlie}} \\ \hline
				$ \ x_0$ & $x_1$ & $\quad \ket{0}\quad$ & $ \quad \ket{1} \quad $ & $\quad \ket{2}\quad$ & $ \quad \ket{0} \quad $ & $\quad \ket{1}\quad$ & $\quad \ket{2} \quad$ \\ \hline
				0 & 0 & 0 & 0 & 0 & 0 & 0 & 0 \\
				0 & 1 & $\frac{2\pi}{3}$ & $\frac{4\pi}{3}$ & 0 & 0 & $\frac{2\pi}{3}$ & $\frac{4\pi}{3}$ \\
				0 & 2 & $\frac{4\pi}{3}$ & $\frac{2\pi}{3}$ & 0 & 0 & $\frac{4\pi}{3}$ & $\frac{2\pi}{3}$ \\
				1 & 0 & $\frac{4\pi}{3}$ & 0 & 0 & 0 & $\frac{2\pi}{3}$ & $\frac{2\pi}{3}$ \\
				1 & 1 & 0 & $\frac{4\pi}{3}$ & 0 & 0 & $\frac{4\pi}{3}$ & 0 \\
				1 & 2 & $\frac{2\pi}{3}$ & $\frac{2\pi}{3}$ & 0 & 0 & 0 & $\frac{4\pi}{3}$ \\
				2 & 0 & $\frac{2\pi}{3}$ & 0 & 0 & 0 & $\frac{4\pi}{3}$ & $\frac{4\pi}{3}$ \\
				2 & 1 & $\frac{4\pi}{3}$ & $\frac{4\pi}{3}$ & 0 & 0 & 0 & $\frac{2\pi}{3}$ \\
				2 & 2 & 0 & $\frac{2\pi}{3}$ & 0 & 0 & $\frac{2\pi}{3}$ & 0 \\		
			\end{tabular}
		\end{center}
		Table S1: Angular phase shifts for secret sharing and DBA experiments. Alice's settings look different compared to Bob's and Charlie's, but they are actually just the same settings multiplied by a global phase. This is due to our particular setup configuration. All these settings belong to mutually unbiased bases for three dimensions. 
		
	\end{table}
	\begin{table}[h!]
		\begin{center}
			\begin{tabular}{c|ccc|ccc}
				\multicolumn{7}{c}{\textbf{Communication Complexity Reduction}} \\ \hline
				\textbf{Settings} & \multicolumn{3}{c|}{\textbf{Alice}} & \multicolumn{3}{c}{\textbf{Bob \& Charlie}} \\ \hline
				$ S $ & $\quad \ket{0}\quad$ & $ \quad \ket{1} \quad $ & $\quad \ket{2}\quad$ & $ \quad \ket{0} \quad $ & $\quad \ket{1}\quad$ & $\quad \ket{2} \quad$ \\ \hline
				0 & 0 & 0 & 0 & 0 & 0 & 0 \\
				1 & $\frac{14\pi}{9}$ & $\frac{16\pi}{9}$ & 0 & 0 & $\frac{2\pi}{9}$ & $\frac{4\pi}{9}$ \\
				2 & $\frac{10\pi}{9}$ & $\frac{14\pi}{9}$ & 0 & 0 & $\frac{4\pi}{9}$ & $\frac{8\pi}{9}$ \\
				3 & $\frac{2\pi}{3}$ & $\frac{4\pi}{3}$ & 0 & 0 & $\frac{2\pi}{3}$ & $\frac{4\pi}{3}$ \\
				4 & $\frac{2\pi}{9}$ & $\frac{10\pi}{9}$ & 0 & 0 & $\frac{8\pi}{9}$ & $\frac{16\pi}{9}$ \\
				5 & $\frac{16\pi}{9}$ & $\frac{8\pi}{9}$ & 0 & 0 & $\frac{10\pi}{9}$ & $\frac{2\pi}{9}$ \\
				6 & $\frac{4\pi}{3}$ & $\frac{2\pi}{3}$ & 0 & 0 & $\frac{4\pi}{3}$ & $\frac{2\pi}{3}$ \\
				7 & $\frac{8\pi}{9}$ & $\frac{4\pi}{9}$ & 0 & 0 & $\frac{14\pi}{9}$ & $\frac{10\pi}{9}$ \\
				8 & $\frac{4\pi}{9}$ & $\frac{2\pi}{9}$ & 0 & 0 & $\frac{16\pi}{9}$ & $\frac{14\pi}{9}$ \\	
			\end{tabular}
		\end{center}
		Table S2: Angular phase shifts for the communication complexity reduction experiment. As in the previous table, Alice's settings are the same as Bob's and Charlie's save for a global phase. 
		
	\end{table}

	\section{Three-arms Interferometer Output Probabilities}
	We write here the output probabilities for a three-arms interferometer.
	\begin{align}
	P(D_0) &= \frac{1}{9} \{ 3+2 \left[ \cos \phi_2 + \cos \phi_3 + \cos (\phi_2 - \phi_3) \right] \} \\
	P(D_1) &= \frac{1}{9} \left\lbrace 3+2 \biggl[ \cos \biggl(\phi_2-\frac{2\pi}{3}\biggr) + \right. \left. \cos \biggl(\phi_3+\frac{2\pi}{3}\biggr) + \cos \biggl(\phi_2-\phi_3+\frac{2\pi}{3}\biggr) \biggr] \right\rbrace \\
	P(D_2) &=  \frac{1}{9} \left\lbrace 3+2 \biggr[ \cos \biggl(\phi_2+\frac{2\pi}{3}\biggr) + \right. \left. \cos \biggl(\phi_3-\frac{2\pi}{3}\biggr) + \cos \biggl(\phi_2-\phi_3-\frac{2\pi}{3}\biggr) \biggr] \right\rbrace \, ,
	\end{align}
	where $\phi_2$ and $\phi_3$ are phase differences relative to the reference arm.
\end{document}